\documentclass[aps,prc,onecolumn,showpacs,superscriptaddress,groupedaddress]{revtex4}

\usepackage{amsmath}
\usepackage{amssymb}
\usepackage{amsfonts}
\usepackage{float}
\usepackage{graphicx}
\usepackage{tabularx}
\usepackage{multirow}
\usepackage{lscape}
\usepackage{rotating}
\usepackage{pstricks}
\usepackage{enumitem}

\begin{document}

\title{A teaching guide of nuclear physics: the concept of bond}

%%%%%%%%%%%%%%%%%%%
%%%%%%%%%%%%%%%%%%%

\author{S. Baroni}
\affiliation{Independent researcher, Barcelona, ES}   

\author{A. Pastore}
\affiliation{Department of Physics, University of York, Heslington, York, Y010 5DD, UK}
     
%%%%%%%%%%%%%%%%%%%
%%%%%%%%%%%%%%%%%%%

\begin{abstract}

We propose discussions and hands-on activities for GCSE and A-level students,
covering a fundamental aspect of nuclear physics: the concept of bond and the energy
released (absorbed) when a bond is created (broken). This is the first of 
the series of papers named \emph{A teaching guide of nuclear physics}, whose main goal is to provide
teaching tools and ideas to GCSE and A-level teachers, within a consistent and complete curriculum.

\end{abstract}

\pacs{ XXX}
 
\date{\today}

\maketitle

%%%%%%%%%%%%%%%%%%%
%%%%%%%%%%%%%%%%%%%

\section{Introduction}
\label{sect:intro}

This work is part of a series of papers where we present activities and discussions that teachers can use to
introduce nuclear physics to GCSE and A-level students, from the basic ideas to the most advanced concepts.
Each aims at guiding teachers and students step by step through
the concepts of nuclear physics.
Using analogies with known situations, hands-on activities
and logics, the students will 
be able to understand fundamental ideas and mechanisms, as
the principle of minimum energy, endothermic and exothermic reactions,
the creation or breaking of bonds.
Stimulating questions that could spark the students' interest
and foster useful discussions are also provided.

As the students progress through the series of papers, these basic concepts 
will allow them to understand a wide variety of phenomena,
as radioactivity, nuclear reactions, nuclear fusion and fission 
\cite{0031-9120-52-2-024001, 0031-9120-52-5-054001}. 

The present paper focuses on the concept of bond and the energy that is released or absorbed when
a bond is created or broken, respectively.

%\section{Prerequisites}
%\label{sect:prerequisites}

%Being this paper the first of the series, there are no prerequisites to understand the material.

\section{Learning objectives}
\label{sect:objectives}

After participating in the activities and discussions described in the next sections,
the students will be able to:
\begin{enumerate}[label=(\alph*)]
\item Describe the concept of bond and explain why
energy is released or absorbed when a bond is formed or broken, respectively.
\item Apply the concept of bond and its associated energy to real-life situations
and explain or predict the outcome of an experiment connecting the macroscopic
observables to the microscopic behaviour of atoms and molecules.
%\item Explain the principle of minimum energy and the concept of stability
%of a physical system and apply these concepts to real-life situations.
%\item Connect the concept of minimum energy with the concept of bond and apply them to
%the atomic nucleus and to radioactivity.
%\item Describe the concept of binding energy of an atomic nucleus and use it
%to explain the valley of stability of the nuclear chart.
%\item Describe the phenomena of nuclear fusion and fission in terms of
%bond creation or destruction and of the binding energy of the involved nuclei.
\end{enumerate}

Along the paper, the symbol \textbf{(Q)} will be used to denote questions for the students.
These questions aims at stimulating further discussions and stimulating the curiosity of students.
The discussions between them and with the teacher are
essential to achieve the overarching goal of these activities: having the students discuss
concepts of nuclear physics and understand them through logics and experiment.
The teacher acts as a facilitator, promoting the good ideas, giving a direction 
to the discussions and encouraging the students to dare making
hypothesis.

We want to stress a very important point here: during the discussions, 
the teacher should promote the usage of a
proper scientific language, with correct terms and verbs throughout the activities.
A correct scientific language is at the basis of the communication among scientists.

As these activity could be conducted with both GCSE and A-level students,
additional material containing more advanced discussions and more 
detailed studies of some concepts are also presented
along the paper. These additional material is denoted as \textbf{(EXTRA)} and can be skipped
when working with GCSE students.

The paper is organized as follows: in section \ref{sect:bonds}, the definition of bond is discussed. 
Section \ref{visual_kit} introduces the visual kit that will be used in
sections \ref{visualize_bonds} to visualize the concept of bond.
Two hands-on experiments will be conducted in sections \ref{hands-on-exp-1} and  \ref{hands-on-exp-2},
where the students will be able to see first hand the effects of creating and breaking bonds at the molecular level.
In sections \ref{evaporation} and \ref{sport_bags}, the concepts that have been learned in the previous sections will be applied to other real-life situations.

\section{Bonds}
\label{sect:bonds}

%\subsection{Definition of bond}\label{definition_bond}

Understanding the concept of (chemical) bond may result complicated for students
\cite{barker2000students, boo1998students}. 
For such a reason, we have developed a simple hands-on activity that may 
take place in the classroom and help fixing basic concepts \cite{brown1992using}.

\textbf{(Q)} What is a bond?

\textbf{(Q)} What does it mean to bond something?

\textbf{(Q)} Can you give some examples of bonds in everyday's life?

The word \emph{bond} has of course many meanings. The ones we are interested
in are well described by the following definitions taken from the 
Cambridge dictionary:

\begin{enumerate}
\item to bond: to stick materials together, especially with glue, or to be stuck together like this.
(example: \emph{This new adhesive can bond metal to glass})
\item bond: a place where single parts of something are joined together, especially with glue,
or the type of join made (examples: \emph{a strong/weak/permanent bond; When the glue has
set, the bond formed is watertight})
\end{enumerate}

\subsection{The visual kit to visualize bonds}\label{visual_kit}

\begin{figure}[!h]
\begin{center}
\includegraphics[width=0.70\textwidth,angle=0]{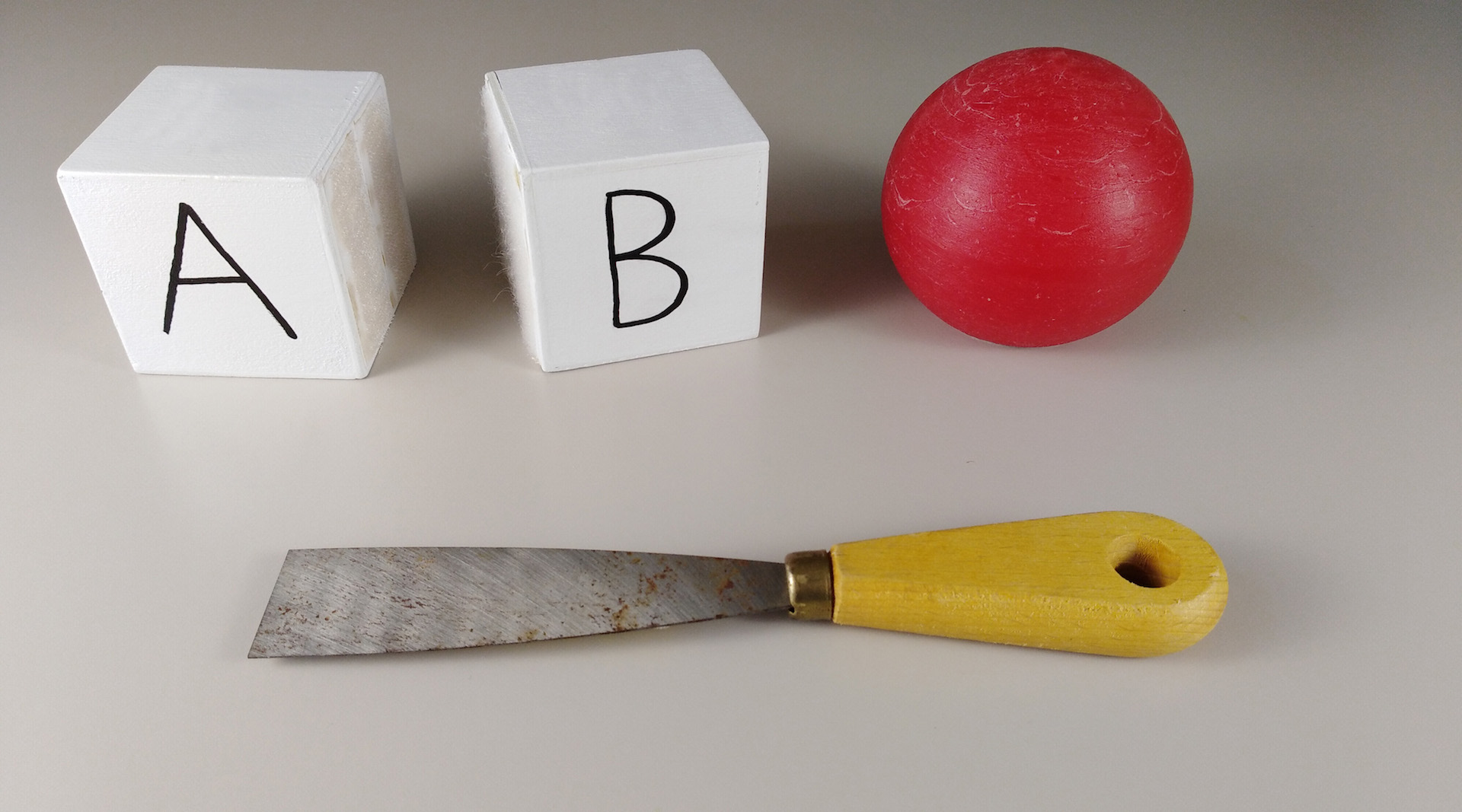}
\end{center}
\caption{(Colors online). The visual kit used to understand the concept of the energy associated to a bond.}
\label{visual_kit_content}
\end{figure}

The following kit (see Fig.\ref{visual_kit_content}) will be used in section \ref{sect:bonds}
to help the students visualize bonds and to understand the concept of the energy associated 
with bonds: 

\begin{enumerate}[label=-]
\item Two cubes labeled with the letters A and B, well visible. Each cube has Velcro on the sides.

\item A sphere: a spherical wax candle would do.

\item A metal or a wooden spurtle.

\end{enumerate}

\subsection{Visualizing the energy of a bond}\label{visualize_bonds}
\textbf{(Q)} What happens when you break a bond?

\textbf{(Q)} What happens when a new bond is formed?

\textbf{(Q)} Think of an example of bond. Now imagine breaking this bond, separating the
two objects that are hold together. What does it take to separate them?

\begin{figure}[!h]
\begin{center}
\includegraphics[width=0.70\textwidth,angle=0]{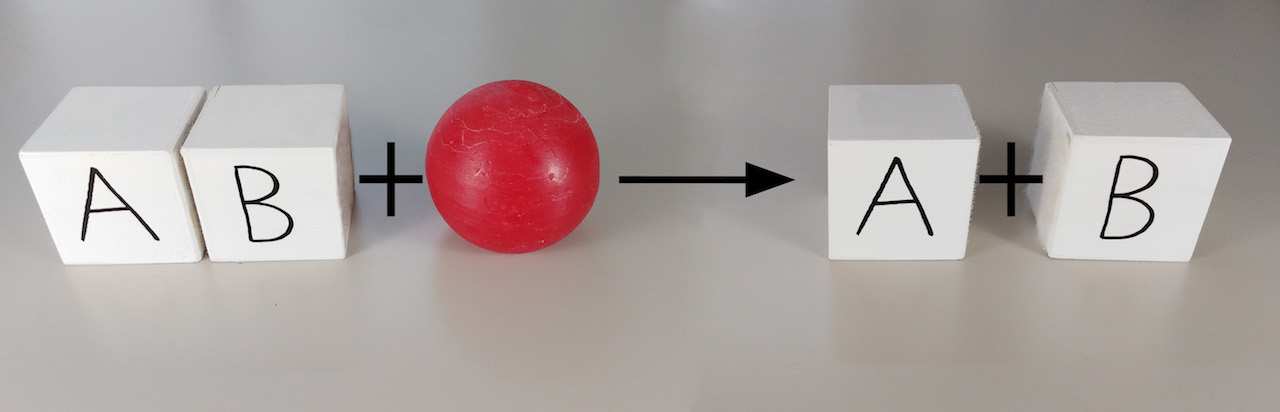}
\end{center}
\caption{(Colors online). The process of breaking a bond, visualized with the kit. The energy (red sphere) is added
to the AB bound system and, as a result, the bond is broken and A and B are separated. Energy has been absorbed
to break a bond.}
\label{visual_kit_equation}
\end{figure}

Help the discussion toward the concept that breaking a bond
requires energy. Once again, the visual kit can be used to visualize this concept:

\begin{enumerate}[label=-]
\item Place the two cubes on the table, joined to each other with Velcro (see Fig. \ref{visual_kit__separating}).
\item Take the spurtle    
and gently start rubbing the spurtle on the Velcro, slowly
separating the two cubes. Explain that the friction between the spurtle, the Velcro and the cubes
is raising the temperature of the system AB.
You could also use an intermediate activity to convince the students of this point:
ask them to rub their hand quickly against the surface of a table. The temperature
of their hand will soon raise, because of friction. 
\item The temperature of the system AB
raises, even if just by a tiny amount, due to the friction with the spurtle.
\item While separating A from B, i.e. breaking the bond, you are transferring
energy to the system AB.

\begin{figure}[!h]
\begin{center}
\includegraphics[width=0.70\textwidth,angle=0]{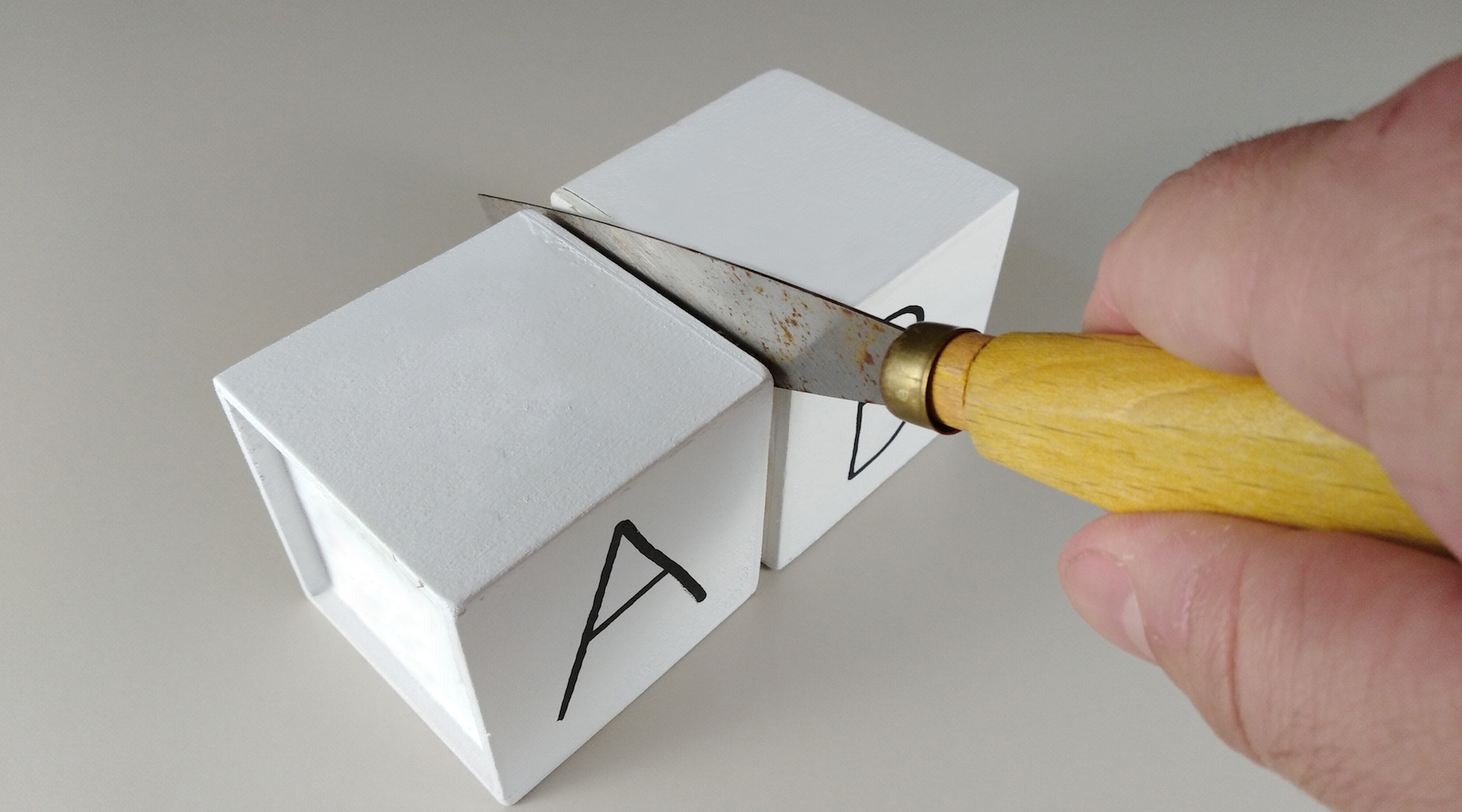}
\end{center}
\caption{(Colors online). Breaking the bond between A and B using a spurtle. The process gives energy to the AB system. This energy is used to break the bond between A and B.}
\label{visual_kit__separating}
\end{figure}

\item Once the two cubes are separated, set the spurtle aside and summarize what happened
using a third object: a sphere representing the energy
that you transferred to the system AB in order to break the bond.
You may label the sphere with the letter E.
Reestablish the bond between A and B, leaving only the cubes A and B on the table.
Then bring in the energy E and put it next to the bond.
The result is that A and B are separated (see Fig. \ref{visual_kit_equation}).
Hence, when the two objects A and B are separated, they have more energy than before breaking the
bond.
\end{enumerate}

\textbf{(Q)} What happens then in the inverse process, when two separated objects are brought
together and a bond is formed?

Let the students play with the visual kit to find an answer.

\begin{figure}[!h]
\begin{center}
\includegraphics[width=0.70\textwidth,angle=0]{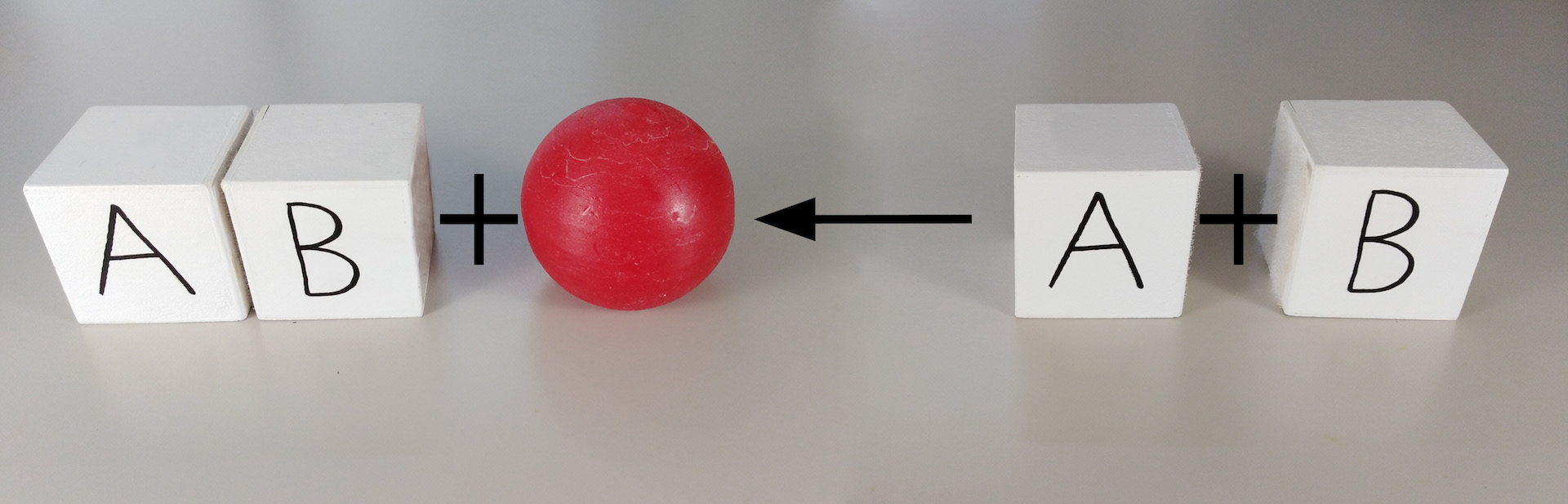}
\end{center}
\caption{(Colors online). The process of creating a bond, visualized with the kit. The A and B components
of the system get bound and energy (red sphere) is released.}
\label{visual_kit_equation_inverse}
\end{figure}

The conclusion that they will soon draw is that:

\begin{enumerate}
\item Energy is absorbed by the system when a bond is broken (process represented in Fig. \ref{visual_kit_equation})
\item Energy is released by a system when a bond is created (process represented in Fig. \ref{visual_kit_equation_inverse})
\end{enumerate}

The reaction that has been represented could be written as follows:

\begin{center}
  AB + energy $\rightleftharpoons$ A + B
\end{center}

where the double arrow indicates that the process can happen in both senses.

Hence, breaking and creating bonds is a way that a system can use
to gain or loose energy. This is a fundamental concept that lies
behind more complex phenomena such as the binding energy of an 
atomic nucleus, the stability of a nucleus, radioactivity, nuclear fusion and fission and many others.
A more detailed discussion on this point can be found in Section \ref{conclusions}. 

\subsection{Hands-on activity: the energy associated with a bond (part I)}\label{hands-on-exp-1}
The goal of this activity is to experimentally observe the energy associated
with bonds. The students will be able to apply the concepts that they learned
in the previous discussions to a concrete case and explain the observed phenomena.

\begin{figure}[!h]
\begin{center}
\includegraphics[width=0.70\textwidth,angle=0]{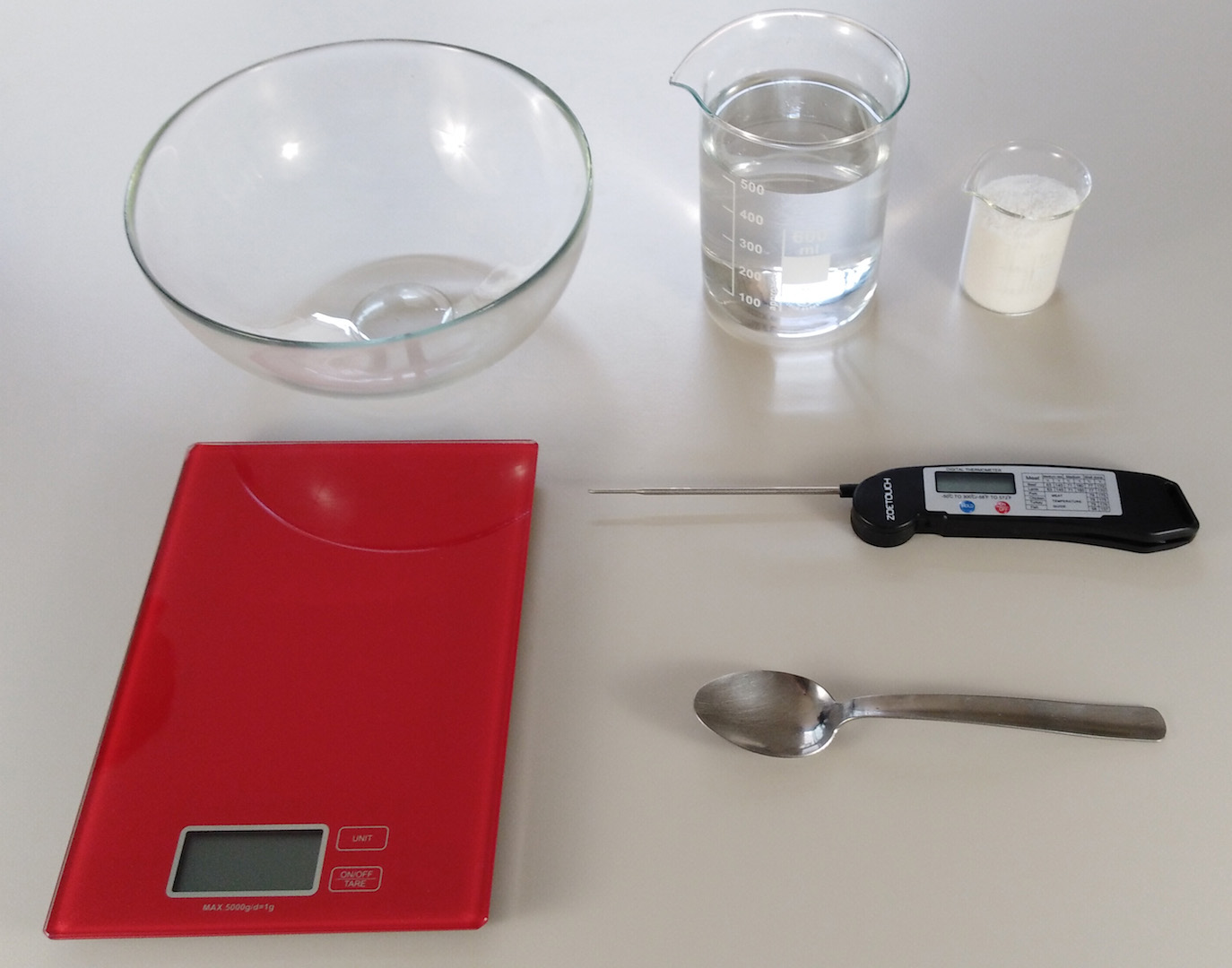}
\end{center}
\caption{(Colors online). The material for the first part of the hands-on activity. From the top left, in clockwise order: a large bowl, a beaker with 500 mL of room-temperature water, about 100 g of kitchen salt (NaCl), a thermometer, a rod or a spoon to stir the water, a scale.}
\label{hands-on_material_1}
\end{figure}

The activity is made of two experiments. The first experiment proceeds as follows:

\begin{enumerate}
\item Split the students into small groups
\item Each group of students is given an experimental kit
\item The kit for this first experiment contains (see Fig. \ref{hands-on_material_1}): a large bowl, a beaker containing 500 mL of room-temperature water (as an alternative any container with the same volume and a scale to measure 500 mL of water), about 100 g of kitchen salt (NaCl), a thermometer (minimum range required: -50 $^{\circ}$C to +50 $^{\circ}$C), a rod or a spoon to stir the water
\footnote{In case that a graduated beaker or container is not available, a scale could be used to weigh the 500 mL of water.}. 
\item Once the students have poured 500 mL of water into the beaker, they have to
measure the temperature of the water (allowing enough time to the
temperature of the water to stabilize after a thermal equilibrium with the container is reached).
\item After that, they have to dissolve salt in the water, stirring the water in order to facilitate the dissolution of the salt
\footnote{WARNING: stirring too quickly could raise the temperature of the system and
alter the results of the experiment by introducing an external source of energy. It is fundamental that one stirs
the water gently.}.
\item It is important that each group measures the temperature at all times and takes notes of what is happening: is the temperature changing? Does all the salt dissolve in the water?
There is no need to explain what they have to focus on: let them experiment with the quantity of salt
that they like and give them enough time to discuss and draw some conclusion.
\end{enumerate}

\begin{figure}[!h]
\begin{center}
\includegraphics[width=0.48\textwidth,angle=0]{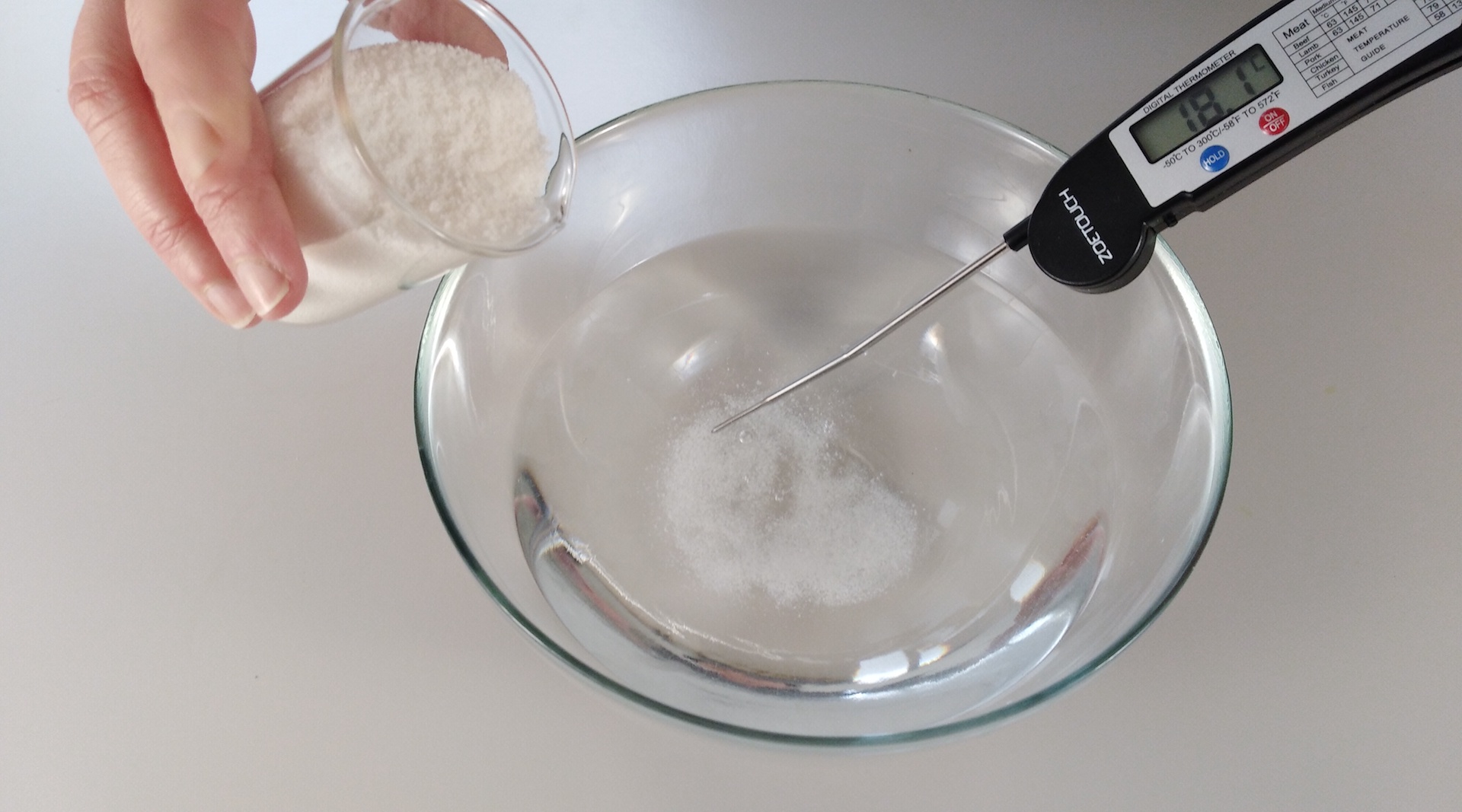}
\includegraphics[width=0.48\textwidth,angle=0]{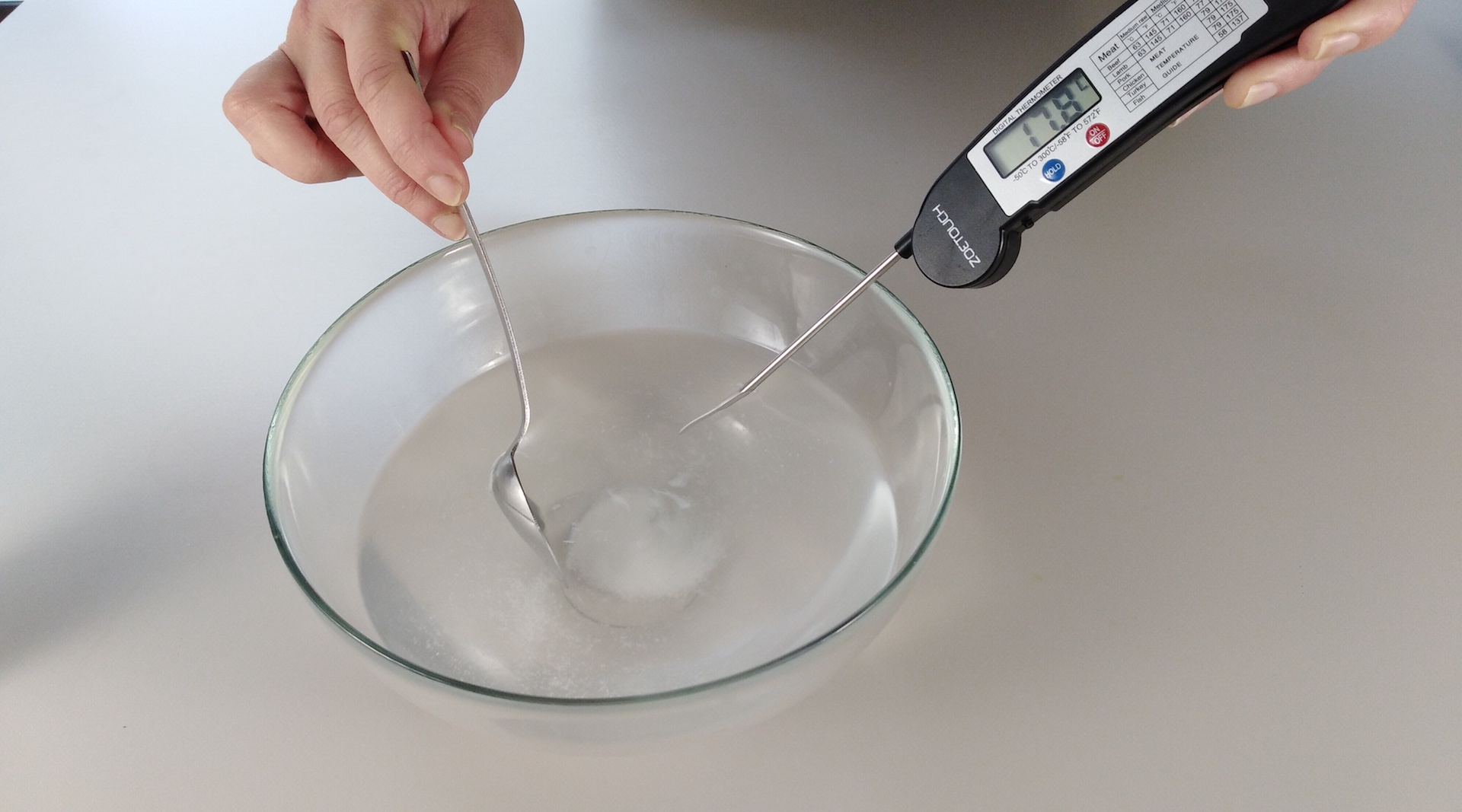}
\end{center}
\caption{(Colors online). The execution of the first part of the hands-on activity. (Photo on the left) Measuring the initial temperature of the water and then starting pouring salt in it. (Photo on the right) Stirring the water with a spoon in order
to dissolve the salt and, at the same time, measuring the temperature at all times. The temperature decreases by
one or two Celsius degrees.}
\label{hands-on_execution_1}
\end{figure}

The temperature of the water typically drops by a couple of Celsius degrees (see Fig. \ref{hands-on_execution_1}).

This first experiment ends with a discussion among all students and the teacher. Guide them through
an analysis of the outcome of the experiment with questions: 

\textbf{(Q)} What happened when you dissolved the salt in the water?

\textbf{(Q)} Why did the temperature drop? How much did it drop? 

\textbf{(Q)} Did it depend on the quantity of salt?

\textbf{(Q)} Why didn't the temperature raise instead of dropping?

\textbf{(Q)} And why did the temperature change in the first place? 

Remind them of the conclusion on the energy associated to a bond and that a system can gain or lose
energy by breaking or creating bonds.

\textbf{(Q)} What bonds could have been created or broken while dissolving the salt in the water?

\textbf{(Q)} If the temperature dropped, does it mean that bonds have been created or broken?

Guide them through the complexity of the process: when a grain of salt is thrown to water,
water molecules surround the grain of salt; bonds are created between water and salt molecules,
while bonds are broken between water molecules to allow the water molecules to join the salt ions;
more and more bonds are created between water molecules and the salt ions
until the water molecules pull salt atoms away from the grain, one by one; bonds are created between
water molecules and salt atoms (Na$^{+}$ cations and Cl$^{-}$ anions) while the bonds between Na and Cl atoms and the bonds between water molecules are broken. 
When the salt is dissolved, there is a huge number of Na$^{+}$ and Cl$^{-}$ ions in the solution, each of them
surrounded by many water molecules. Each bond between a water molecule and a salt ion is weak, but the
number of bonds is large.

\textbf{(EXTRA)} If the students are interested and if their level of chemistry allows it,
the discussion can temporarily divert to the kind of bonds that keep the salt ions together (ionic bonds)
and the kind of bonds that water molecules can form with each other and with the salt ions 
(hydrogen bonds) \cite{klein2017organic}.

\textbf{(Q)} Some bonds are created, then, while some other bonds are broken. Which one is larger: the energy released
by the bonds that are created or the energy that is absorbed by the bonds that are broken? How can you tell?

\textbf{(Q)} How could we make this effect larger? How could we make the temperature drop much more dramatically?

The stronger the broken bonds are, the more the energy they absorb and the more the temperature drops.
The bonds that were broken in the first experiment were: the bonds between water molecules and the bonds
between the salt ions. We can not do much about the bonds between the salt ions, but we can surely make
the bonds between water molecules much stronger: we could use ice instead of water.

\textbf{(EXTRA)} Energy is absorbed to break the bonds and this makes the temperature drop. The link between the two is 
given by the atomic theory and the kinetic theory of gases \cite{boltzmann1896gases, kauzmann1914kinetic}. 
The energy to break the bonds is taken from the kinetic energy
of the surrounding molecules, which then slow down. Slower molecules correspond, on a macroscopic point of view, to a lower temperature.

\subsection{Hands-on activity: the energy associated with a bond (part II)}\label{hands-on-exp-2}

\begin{figure}[!h]
\begin{center}
\includegraphics[width=0.70\textwidth,angle=0]{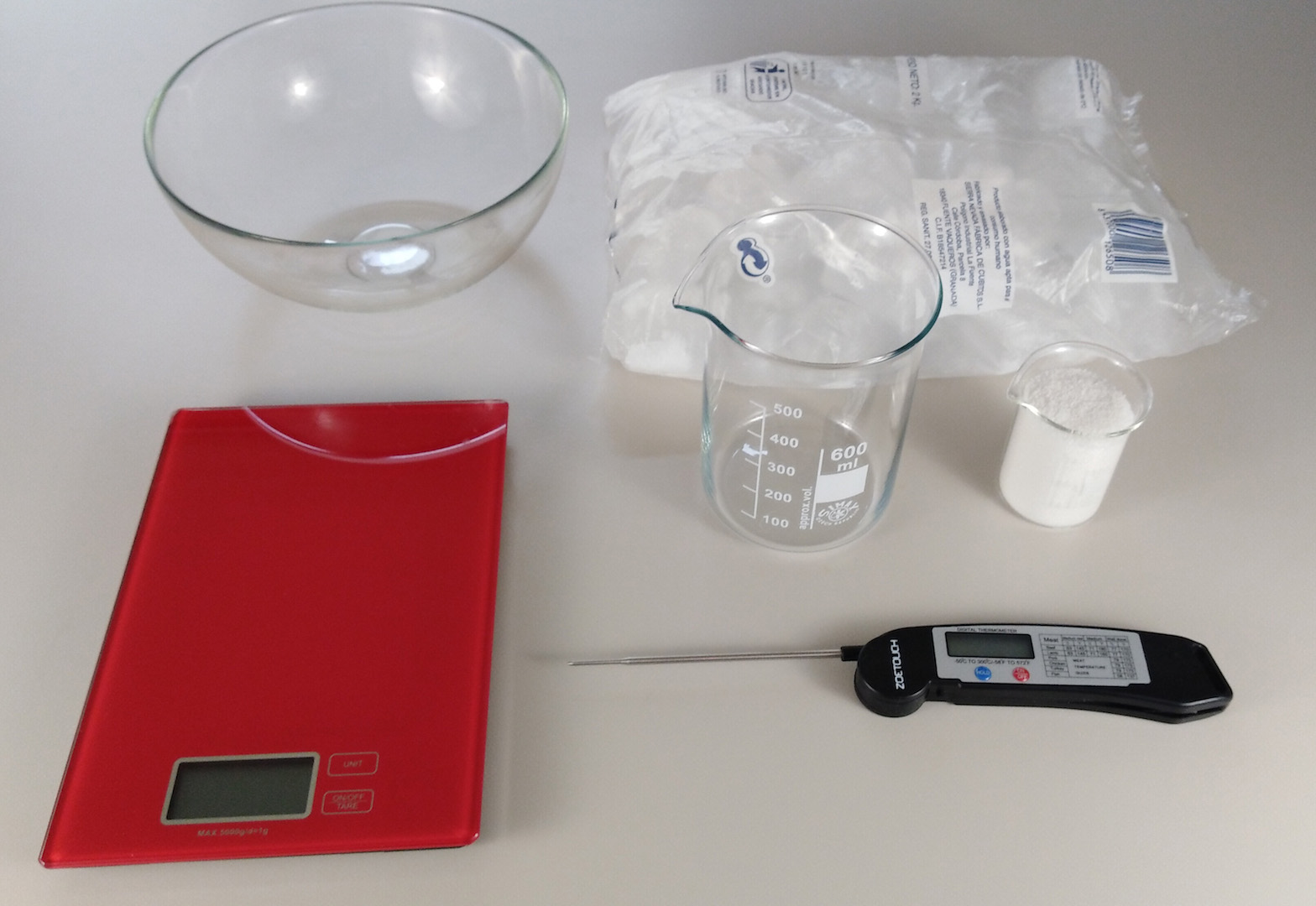}
\end{center}
\caption{(Colors online). The material for the second part of the hands-on activity. From the top left, in clockwise order: a large bowl, a bag of ground ice (at least 400 g), a beaker with 500 mL of room-temperature water, about 100 g of kitchen salt (NaCl), a thermometer, a rod or a spoon to stir the water, a scale.}
\label{hands-on_material_2}
\end{figure}

The second experiment proceeds as follows:
\begin{enumerate}
\item Keep the same groups of students as in the first experiment
\item Each group of students is given an experiment kit
\item The kit for this second experiment contains (see Fig. \ref{hands-on_material_2}): a large bowl, a bag of ground ice (at least 400 g), a beaker containing 500 mL of room-temperature water (or, as an alternative, any container with the same volume and a scale to weigh 500 mL of water), about 100 g of kitchen salt (NaCl), a thermometer (minimum range required: -50 $^{\circ}$C to +50 $^{\circ}$C), a rod or a spoon to stir the water
\item The students have to pour 500 mL of water into the beaker and add about 10 cubes of ice. It is important
that all groups add approximately the same amount of ice, so that possible differences cannot be associated
to a different setup of the experiment.
\item The students have to measure the temperature of the water+ice (wait for the
temperature to stabilize after a thermal equilibrium with the container has been reached)
\item After a thermal equilibrium is reached, they can dissolve the salt in the water, stirring the water to
facilitate the dissolution of the salt. 
%\footnote{WARNING: stirring too quickly could raise the temperature of the system and
%alter the results of the experiment by introducing an external source of energy. It is fundamental that one stirs
%the water gently.}.
%\footnotemark[1]
\item It is important that each group measures the temperature at all times and takes notes of what is happening: is the temperature changing? Does all the salt dissolve in the water? Let them experiment with the quantity of salt
that they like and give them enough time to discuss and draw some conclusion.
\item Once the temperature seems not to drop anymore, let the students add more ice. Let them experiment freely. You could even transform this part of the experiment into a competition: which group will reach the lowest temperature? 
\end{enumerate}

\begin{figure}[!h]
\begin{center}
\includegraphics[width=0.48\textwidth,angle=0]{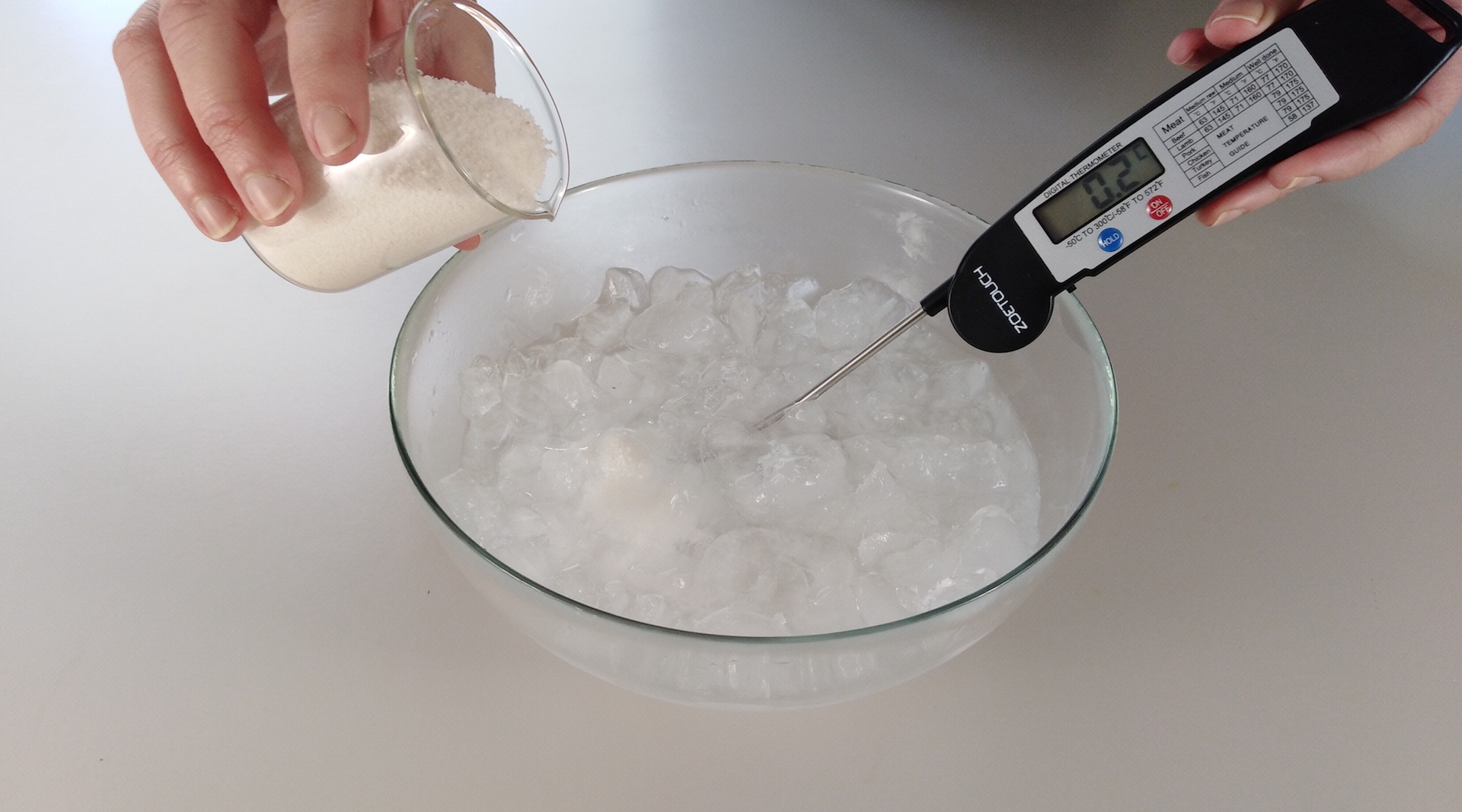}
\includegraphics[width=0.48\textwidth,angle=0]{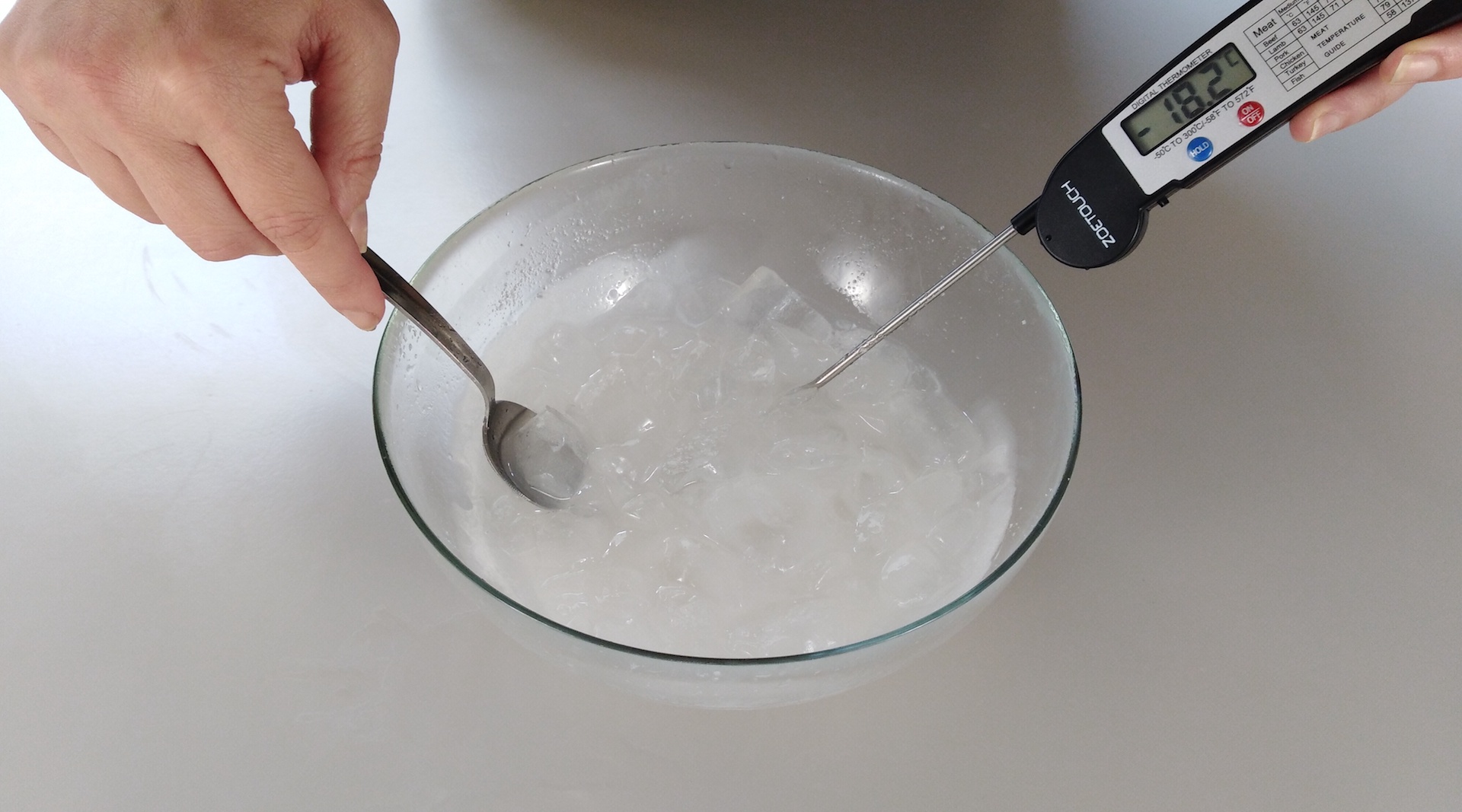}
\end{center}
\caption{(Colors online). The execution of the second part of the hands-on activity. (Photo on the left) Measuring the initial temperature of the water+ice system (around 0 degrees centigrades) and then starting pouring salt in it. (Photo on the right) Stirring the solution with a spoon in order
to dissolve the salt and, at the same time, measuring the temperature at all times. The temperature decreases very quickly and reaches about -18 $^{\circ}$C.}
\label{hands-on_execution_2}
\end{figure}

Before adding salt, the temperature of the water+ice system is about zero degrees centigrade, which 
is the melting temperature of the ice. After adding the salt, the temperature
gradually drops to -18 $^{\circ}$C (see Fig. \ref{hands-on_execution_2}). The more salt is dissolved, the lower the temperature gets.

\textbf{(Q)} How can you explain what happened in terms of bond creation or destruction?

Compared to the first experiment, there is an additional bond that is broken: the bond between water molecules in the ice. As this bond is much stronger than the bond between water molecules in water, the energy absorbed when this bond is broken
is much higher. Hence the temperature drops much more with respect with the first experiment.
This also explains why the temperature of the water+ice+salt system does not drop so strongly if the ice
is floating on the surface of the water, without getting in contact with the salt on the bottom of the container.
The most dramatic drop of temperature is attained if the ice cubes touch the salt on the bottom, as this causes
a large number of bonds to be broken between water molecules in the ice. Hence, a good suggestion to the students, in
case they were not able to reach the -18 $^{\circ}$C, would be to get rid of some of the water, so that the ice cubes
could get in contact with the salt.

\subsection{Hands-on activity: evaporation of alcohol}\label{evaporation}
In this activity, the students will be able to apply what they learned in the previous sections
to the case of the evaporation of a liquid. 

\begin{enumerate}
\item Keep the same groups of students as in the previous experiments
\item Give a bottle of room-temperature denatured alcohol to each group and ask the student to pour
some drops of alcohol on the back of their hands and to wait for a couple of minutes.
\item Ask them to be patient and to register any change that they may feel.
\end{enumerate}

\textbf{(Q)} What happened after you poured some alcohol on your skin?

\textbf{(Q)} Did it feel hotter or colder?

\textbf{(Q)} Write down a possible explanation for the fact that the interested area
felt colder after some time.

\textbf{(Q)} What does it mean, at the molecular level, that the alcohol evaporated?

The molecules of denatured alcohol are short carbon chains made of only two carbon atoms
(CH3-CH2-OH) \cite{klein2017organic_b}. This facilitates their evaporation. 
Evaporation means that the bonds between the alcohol molecules at the surface of the liquid are broken
and these molecules can then abandon the liquid and enter the gas phase of the surrounding air.
To break the bonds between the alcohol molecules in the liquid phase, energy is needed, as we
learned in the previous sections. This energy is taken from the kinetic energy of the molecules 
that the skin is made of and this in turn corresponds, at a macroscopic level, to a lower temperature.

\subsection{Hands-on activity: sport bags}\label{sport_bags}
In this activity, the students will be able to apply what they learned in the previous sections
to the case of the well-known bags that are used in sports to cool down or heat up very quickly an athlete's body.

\begin{enumerate}
\item Keep the same groups of students as in the previous experiments
\item Give one bag to each group. Some of the bags, when shacked, will cool down quickly. Some others
will heat up.
\item Tell the students that they cannot open the bags, but they can shake them or fold them. Let them experiment
freely until they discover that the temperature of the bag raises or drops after shaking it.
\end{enumerate}

\textbf{(Q)} What happened with your bag? Did the same happen for all groups?

\textbf{(Q)} Write down a possible explanation for what happened with your bag.

\textbf{(Q)} Write down a possible explanation for the fact that other groups' bag experienced
the opposite change of temperature.

After the discussion, have the different groups compare their explanation and agree on a common conclusion.
The difference between the two kind of bags is that some salts allow a large number of water molecules
to surround each salt ion. This results in a larger number of bonds that are formed and hence 
a larger energy that is released. If the energy released is larger than the energy that is required
to break the water-water and ion-ion bonds, then the system's temperature raises.
With some other salts, the number of water molecules surrounding each ion is not large enough
and the energy released is less than the energy required to break the water-water and ion-ion bonds.
Hence the temperature of the system drops.

%%%%%%%%%%%%%%%%%%%
%%%%%%%%%%%%%%%%%%%
\section{Conclusions}\label{conclusions}

It is fundamental to help the students to draw conclusions at the end of the activities and to summarize
what has been achieved. This helps the students to fix the ideas and to put them into context.

The most effective way to summarize the activity is doing it interactively. Once again, the teacher acts 
as a facilitator, promoting the good ideas and giving a direction to the discussions.

\textbf{(Q)} What fundamental concept have we touched upon in these activities?

\textbf{(Q)} Why is the concept of bond so important?

Guide the students toward the idea that creating and breaking bonds allows a system to exchange
energy with the exterior, and that energy exchanges are essential for a system to evolve. Without entering
into the details, you could mention the principle of minimum energy that states that for a closed system
the internal energy will tend to decrease and to approach a minimum at equilibrium. A closed system is a system that can exchange energy with the exterior, but not matter. The principle of minimum energy, which will be covered in
the following papers of this series, implies that a system naturally evolves toward a minimum energy configuration
through energy exchanges with the exterior.

%As an example, you can briefly talk about radioactivity \cite{krane1988introductory}:
%
%\textbf{(Q)} What is radioactivity?
%
%\textbf{(Q)} What does it mean that a nucleus is radioactive?
%
%\textbf{(Q)} Why does a radioactive $^{238}U$ atom naturally emits an alpha particle?
%
%An alpha particle is a nucleus of ${}^{4}He$ and it is made of 2 neutrons and 2 protons.
%The radioactive decay process for the $^{238}U$ atom reads:
%
%\begin{center}
%	$^{238}U \rightarrow  {}^{234}Th +  {}^{4}He$ + energy
%\end{center}
%
%Splitting the $^{238}U$ atom requires energy, as bonds between nucleons have to be broken, but
%at the same time the bonds inside $^{234}Th$ and  $^{4}He$ are stronger than those in $^{238}U$.
%Hence stronger bonds are created with respect to $^{238}U$. The energy released is larger than the
%energy needed to split the $^{238}U$ atom in two parts, hence energy is released in the process and the
%two atoms of $^{234}Th$ and  $^{4}He$ have a lower energy than the initial $^{238}U$ atom.
%
%Of course the effect of creating and breaking bonds does not only affect the atomic level, but it can also be perceived at the macroscopic level. 

We saw examples where the effect of a huge number of bonds
leads to a macroscopic increase or decrease of the temperature of the system. The students can be guided through
this line of reasoning with questions as:

\textbf{(Q)} Why did the temperature drop when we added salt to the water with ice?

\textbf{(Q)} What energy transformations are taking place in the process?

The total energy absorbed by the broken bonds is larger than that released by the creation of new bonds. This energy 
is taken from the kinetic energy of the surrounding molecules and this is perceived as a drop in temperature.
%between molecules, atoms and nucleons
In conclusion, have the students bring a clear message home: the bonds  and their associated energy are two fundamental
concepts in physics, and as the activities and the discussions in the following papers of this series will show,
these concepts are at the basis of many important phenomena in nuclear physics.
%%%%%%%%%%%%%%%%%%%
%%%%%%%%%%%%%%%%%%%
\section*{Acknowledgments}
A.P. acknowledges a support from the STFC Grant No. ST/P003885/1 and ST/P006213/1.

%%%%%%%%%%%%%%%%%%%%%%%%%%%%%%%%%%%%%%%%%%%%%%%%%%%%%%%%%%%%%%%%%%%%%%%%%%%%%%%
%%%%%%%%%%%%%%%%%%%%%%%%%%%%%%%%%%%%%%%%%%%%%%%%%%%%%%%%%%%%%%%%%%%%%%%%%%%%%%%

\bibliography{biblio}

\end{document}